# Predicted reentrant melting of dense hydrogen at ultra-high pressures

Hua Y. Geng* and Q. Wu

*National Key Laboratory of Shock Wave and Detonation Physics, Institute of Fluid Physics, CAEP; P.O.Box 919-102, Mianyang, Sichuan, P. R. China, 621900*

**The phase diagram of hydrogen is one of the most important challenges in high-pressure physics and astrophysics. Especially, the melting of dense hydrogen is complicated by dimer dissociation, metallization and nuclear quantum effect of protons, which together lead to a *cold melting* of dense hydrogen when above 500 GPa. Nonetheless, the variation of the melting curve at higher pressures is virtually uncharted. Here we report that using *ab initio* molecular dynamics and path integral simulations based on density functional theory, a new atomic phase is discovered, which gives an uplifting melting curve of dense hydrogen when beyond 2 TPa, and results in a reentrant solid-liquid transition before entering the Wigner crystalline phase of protons. The findings greatly extend the phase diagram of dense hydrogen, and put metallic hydrogen into the group of alkali metals, with its melting curve closely resembling those of lithium and sodium.**

*Corresponding author: s102genghy@caep.cn

Under compression conditions, hydrogen and alkali metals have unfolded exotic and fascinating properties, and have attracted broad attention and interests[1-6]. Though hydrogen is assigned to the IA group in the periodic table, at low pressure it is similar to halogens rather than alkalis, with a high electronegativity and a tendency to form covalent bond with itself. At ambient conditions it exists as diatomic molecular gas, instead of in a BCC or FCC crystalline structure that usually adopted by simple metals[7]. However, recent high-pressure experiments and theoretical predictions have shed new light on their distinctions, and hinted interesting similarities and parallels for the compression behavior of dense hydrogen and light alkalis.

For example, under compression both H and Li undergo an *s-p* electronic transition, which leads to complex structure with low coordination numbers[8-10]. In particular, Li, Na, and H all exhibit a maximum in their melting curves, as well as the subsequent *low-temperature* melting[2-4,11]. It was speculated that except some details, the phase diagram of hydrogen and alkalis should be qualitatively similar. This paralleling, however, is questioned when going to higher pressures, where hydrogen was predicted to continuously lower its melting curve and enter an *ultra-cold* quantum liquid[1,12,13], whereas both experiments and theoretical calculations indicated a follow-up increase of the melting temperature for Li and Na. On the other hand, a recent deliberate calculation[2] suggested a flat melting curve for dense hydrogen between 500 GPa and 1.5 TPa. This variation resembles the melting minimum observed in both Li and Na[3,4]. We will strengthen this parallelism below, by unveiling a new atomic phase of dense hydrogen and the concomitant ascent of the melting curve when beyond 2 TPa, with the melting temperature $T_m$ reaches ~700 K at 7 TPa. This discovery greatly enhances the occurrence probability of solid metallic hydrogen in the core of *cold* giant gas planets.

The computed melting curve is drawn as the filled square-solid line in Fig. 1. We found that the $T_m$ of H is flat between 500 to 1500 GPa, as demonstrated in Ref. [2]. It however increases rapidly as compression goes further and transition into a new atomic phase. This ascent is intriguing since a zero Kelvin quantum liquid of dense hydrogen had been hypothesized in this pressure range[1,12,13]. Analysis shows that the melting minimum created by this reentrant transition, which spans from 500 GPa to 1.5 TPa, is a direct consequence of the competition between chemical covalent bonding (that favors *quasi*-molecular motifs) and pressure-induced orbital delocalization (that prefers metallization)[14]. The resultant frustration greatly weakens the crystalline ordering. It lowers the melting curve to a minimal level, which coincides exactly with the $H_2$ dissociation line, as illustrated in Fig. 1. This mechanism naturally interprets the melting maximum presented in the molecular phase of $H_2$ at ~100 GPa: the compression-driving evolution towards $H_2$ dissociation, or the gradually increased occupation of electrons in the $\sigma^*$ anti-bonds and *p* orbitals, weakens the crystalline stability[15]. Beyond this reentrant melting, figure 1 indicates an energy gain by crystalline ordering from the metallic interaction that prevailing at TPa regime.

Similar phenomenon has been observed in Li and Na, where the melting minimum is created mainly because of *s*-*p*(-*d*) transition[3,4]. There are no molecular phases involved in dense Li and Na, but strong competition between chemical bonding and metallic interaction still presents, as the tendency towards pairing and formation of low-symmetric structures with low-coordination numbers that frequently observed in dense Li and Na unequivocally illustrated[8-10]. Analogously, the wide flat range of $T_m$ beyond $H_2$ dissociation as shown in Fig. 1 is also considered as due to excitation of some electrons to orbitals with *p* character, as implied by the $H_3$ units and chain structures prevailing in many (meta-)stable solid structures predicted in this pressure range[5,14].

In particular, our results show that the previously proposed groundstate[16] *Cmcm*-8 (a structure has the *Cmcm* space group symmetry with 8 atoms in the primitive cell) is thermodynamically unstable with respect to a newly discovered $C222_1$-8 phase. The number of atoms in the primitive cell will not be given explicitly below for the sake of brevity. For example, *Cmcm*-8 will be abbreviated to *Cmcm* if there is no confusion. The discovery of a new stable structure by MD simulations signals the challenge in structure prediction based solely on the static lattice enthalpy and harmonic phonons. The latter explores only a small area surrounding the ideal atomic positions, thus incapable of describing the lattice dynamics of dense hydrogen reliably[14]. In terms of static lattice enthalpy, the phases of *Cmcm* and $C222_1$ are almost degenerated, as demonstrated in Fig. 2(a), in which the enthalpies are represented as a difference with respect to a predefined equation of state. The $C222_1$ structure has an enthalpy slightly lower than that of *Cmcm.* But the difference is less than 1.5 *m*eV/H. When classical nuclear motions are included by using AIMD simulations, however, the *Cmcm* phase spontaneously collapses into $C222_1$ phase at an equilibrated temperature as low as 20 K. This transition keeps the orthorhombic lattice unchanged but the symmetry is lowered by local atomic displacements, implying that dense hydrogen still disfavors highly symmetric structures even at such high pressures. The electron localization function (ELF) as shown in Fig. 2(a) clearly unveils that the stability of this phase closely relates to the competition between residual chemical bonding and renascent metallic interactions: the *Cmcm* phase shows obvious $H_2$ bonding within the $H_3$ unit, whereas such pairing does not present in the $C222_1$ structure. Namely, it is the distortion away from this kind of *quasi*-molecular motifs that lowers the structural symmetry[5,14,16].

It should be pointed out that such *quasi*-molecular units also exist at low pressures down to 500 GPa. But in that pressure range they are meta-stable and the groundstate is the degenerated atomic phases of $I4_1/amd$-2 (which is also the Cs-IV phase) and/or *Fddd*-2 (a distortion of the Cs-IV phase)[14]. Strong compression increases the occupation of electrons on the anti-bonding $\sigma^*$ states[17], some electronic orbitals even have *p* character. One consequence of this effect is the creation of the melting minimum and the dissociation of hydrogen from molecular state into atomic state at around 400~500 GPa[5,17,18]. The reappearance of *quasi*-molecular units in the *Cmcm* and $C222_1$ phases at ultra-high pressure of terapascals therefore cannot be attributed to enhancement of the chemical interactions by compression. One interpretation is that compression brings protons close enough (to ~0.84 Å) so that the continuously *diminishing* chemical bonding can still play its role to some extent[17]. We indeed find that such bonding is very weak and can be easily destroyed by nuclear thermo-motions.

The key structural feature that differentiates phase $C222_1$ from *Cmcm* is that the former does not have a clear-cut coordination number gap as the latter at a radius between 0.94 and 1.24 Å. Therefore, atoms in the $C222_1$ phase can access a larger configurational space than in the *Cmcm* phase, and thus has greater entropy. The lack of a coordination gap also implies that $C222_1$ is more *metallic* than the *Cmcm* phase. Dense hydrogen is obviously prone to this atomic structure when nuclear kinetics is taken into account, and the $H_3$ units might be just transient clusters, as suggested by the radial distribution function (RDF) of this structure at finite temperatures shown in the inset of Fig. 2(b). At 50 K, the first coordination shell of $C222_1$ splits into two peaks, with the first sub-shell contains 6.7 protons in average and 5.3 protons in the second one. At higher temperatures, however, these peaks merge into a single one, with 12 protons as the nearest neighbors, the same coordination number as the close-packed simple metals[7].

The electronic structure of these phases corroborates above observations exactly. As can be seen from Fig. 3, they are very similar to the nearly free electron gas model for simple metals, which is particularly evident for the FCC structure. Both *Cmcm* and $C222_1$ phase show similar features: In the vicinity of the Fermi level a pseudo-gap is created. This dip in the density of states could be ascribed to charge density wave, a mechanism commonly observed in alkalis[7]. Alternatively, it can approximately be interpreted as due to the formation of weak chemical bonds in real space, which is helpful for understanding the *quasi*-molecular $H_3$ unit. The band structure plotted in Fig. 3 shows that *p* orbitals take an important role in the stability of these phases. All of the states near the Fermi level show pronounced *p* character, and it is the splitting of the states along the **S-R** direction in the reciprocal space that opens the pseudo-gap up, as a result of strong hybridization between *s* and *p* orbitals. In other words, it is the compression that brings nearest-neighboring hydrogen atoms to a distance close enough and the subsequent *sp* hybridization that stabilizes $C222_1$ and *Cmcm* phases.

Though *Cmcm* and $C222_1$ have similar electronic structure at the static lattice approximation, different from *Cmcm*, the dynamical stability of $C222_1$ structure is quite robust. It possesses the highest melting temperature and the highest superheating limit from 2 to 5 TPa among all studied structures (Fig. 4). The cI16 phase has the second highest melting curve (not shown). Beyond 5.3 TPa the high-symmetric FCC becomes the most stable solid phase with the highest melting temperature and the highest superheating limit, followed by BCC as the second stable one. It is interesting to observe that $I4_1/amd$ phase reaches a melting minimum at about 3 TPa (where it becomes metastable), suggesting that the stabilization of this phase at lower pressures[5,14] is actually resultant from interaction frustration. It also should be noted that at around 4 TPa, the FCC phase can be stabilized by a pure exponentially decayed repulsive potential, demonstrating that the dense hydrogen already entering the Wigner crystal region[19].

Because the Z-method is prone to overestimating the $T_m$ in some systems[2], in order to verify its validity, we also calculate $T_m$ by using the two-phase method with *NPT* ensemble (TPM-NPT)[20]. At 2 and 4 TPa, the obtained $T_m$ by TPM-NPT method perfectly overlap with the results of Z-method (Fig. 4), justifying that Z-method works well in dense hydrogen[2]. The application of PBE functional to metallic hydrogen with quasi-molecular units might also raise some concerns, since it is well-known that $H_2$ dissociation is poorly described by this functional. Nonetheless, considering the nearly free electronic features as revealed in Fig. 3, it is plausible to expect that PBE performs well in this circumstance. This argument is justified by using the van der Waals functional vdW-DF with TPM-NPT method, as shown in Fig. 4. This van der Waals functional has been carefully examined recently and was shown of good performance in $H_2$ dissociation by comparison to both accurate QMC simulation[21] and dynamic compression experiment[22]. The good agreement of the $T_m$ calculated by vdW-DF and PBE functionals unequivocally demonstrates the validity of PBE approximation at such high pressures. The NQE on the phase stability and superheating limit is assessed by using *ab initio* path integral MD (AI-PIMD)[2,19]. The conclusion is that the NQE correction is very small, as demonstrated by the superheating limit $T_{sl}$ of the *C222₁* phase in Fig. 4. This is drastically different from the *Fddd* and Cs-IV phases at pressures between 500 and 1500 GPa, where a pronounced NQE was observed[1,2]. It is worth mentioning that this does not imply that the absolute contribution of NQE to the free energy is small. Rather, it simply indicates a cancellation of the NQE contribution between the liquid and solid phases. Similar cancellation was also observed in the melting of Li[23]. These observations consolidate and validate the melting curve obtained with Z-method.

One of the main findings is that beyond 3 TPa, all considered candidate solid phases enhance their stability with increasing pressure, which is reflected in the ascending $T_m$ and $T_{sl}$. Our work reveals that this increase of the stability is dominated by the potential energy surface. NQE has little influence on this respect. This discovery might be helpful for searching for the quantum liquid of metallic hydrogen, which was speculated as a superconducting superfluid[24] at a pressure range of 0.5~1.5 TPa, where a metastable supercooled liquid state might exist[2]. When beyond 7 TPa, our result indicates that nuclear-nuclear repulsion takes the main role in determining the atomic structure, and the electronic structure can be described by a free electron gas model. Therefore, low-symmetric structures such as those presented between 0.2 to 5 TPa are unlikely any more. On the other hand, theoretical analysis suggested that in Wigner crystalline phases, the BCC is slightly stable than FCC phase, thus there might still have a FCC→BCC transition at higher pressures. Beyond that there would have no other solid-solid phase transition can be expected, except the nontrivial melting of Wigner crystals induced by enormous nuclear quantum kinetic energy.

The finally obtained phase diagram is summarized and plotted in Fig. 1, in which the vertical dash-dotted lines that separate atomic solid phases are roughly based on the intersections of the respective melting curves, and are guides to the eye only. Reliable experimental data about the melting of dense hydrogen[11,25-27] available so far are also included for comparison (here we do not distinguish the data made on deuterium from those on hydrogen, since at high temperatures the isotopic effect is small and they become similar). Especially, the recent Raman data, which did not measure the molecular solid phase melting directly but gave a hint about the lower bound of the melting curve, suggested that the melting temperature of dense molecular hydrogen indeed decreases with increasing pressure, and follows the variation trend predicted by the Kechin's equation up to 250 GPa[11]. These experimental data are in good agreement with the theoretical predictions given in Refs. [2,28,29]. This, from another point of view, is a direct experimental validation of the theoretical method we employed. A triple point between phases I-IV-liquid was also hinted in the Raman spectroscopy experiment[11]. It implied that the slope of the melting curve could increase slightly beyond that triple point. The theoretical data of the melting curve around this pressure do hint such a slight change. This comparison also implies other two triple points between phases IV, V, and the molecular liquid, and that between phase V, the molecular liquid, and the atomic liquid. It is worth pointing out that both the dissociation line predicted by Ref. [30] and the recent dynamic experimental data by Ref. [22] match very well with each other, and all point to the coalescence with the melting curve at about 490 K and 390 GPa. This dissociation line separates the semiconducting and metallic states in the liquid phase, and might extend to the solid regime to separate the phase V and the atomic $I4_1/amd$ phase[6], as the dashed line in Fig. 1 indicated.

This diagram outlines the overall structure of the phase diagram of dense hydrogen within the pressure range that will be accessible by dynamical compression in the near future[31]. It greatly advances our understanding about dense hydrogen, and unambiguously demonstrates that the cold melting of dense hydrogen directly relates to the dissociation of $H_2$. Namely, it is resultant from the flat potential energy surface due to the occupation of electrons on both the bonding and anti-bonding states, which leads to a strong competition and frustration between electron localization and delocalization. In addition, most gas planets contain hydrogen as the main composition with the core pressure up to terapascal scale. Our findings pave the way to evaluate the high-pressure properties of hydrogen-rich minerals and rocks that are crucial for modelling the interior structure and dynamical evolution of theses planets. The computation and analysis techniques employed here could also be applied to assess the phase diagram of other substances under extreme conditions.

## Methods

**Electronic structure calculations.** The density functional theory is used to describe the electronic structures. The interaction between electrons and protons is described by all-electron projector augmented wave (PAW) method, and the Perdew-Burke-Ernzerhof (PBE) electronic exchange-correlation functional is employed. Both the standard and the hard version of the PAW pseudopotentials provided in VASP code are used. The validity of these potentials is carefully checked against the all-electron full-potential results of WIEN2k, as prescribed in Ref. [2]. An energy cutoff of 700 eV is used for the plane-wave expansion of the valence and conduction band wave functions. A $3\times3\times3$ Monkhorst-Pack *k*-point mesh is used to sample the first Brillouin zone. A larger $4\times4\times4$ mesh is also used to check the convergence quality. The size of the simulation cell varies from containing 432 to 500 hydrogen atoms depending on the candidate solid structures, with periodic boundary conditions being imposed.

**Melting curve calculations.** The melting curve of dense hydrogen is modeled using *ab initio* molecular dynamics (AIMD) simulations. In order to obtain a reliable melting curve, the ground-state candidate structures that are predicted to be stable at this ultra-high pressure range and all other known low-lying structures are explored, including the *Fddd* and Cs-IV phases[14], hexagonal and planar $R\bar{3}m$, triatomic *R3m* and *Cmcm* phases, the high-symmetric BCC and FCC, and cI16-$I\bar{4}3d$ phases[5], *etc.* The cI16 is a distortion of the BCC structure and is also the high-pressure phase reported for Li and Na when the slope of their melting curve becomes positive again[9]. It is necessary to point out that it is impossible to consider all possible structures. The physical melting temperature always corresponds to that of the structure that has the highest $T_m$, and what we obtained here is an estimate of the lower bound of the melting curve. The true $T_m$ might be higher if structures more stable than we employed here were found. However, this will not modify our conclusion that the melting curve of metallic hydrogen has a positive slope when beyond 2 TPa. The Z-method that is widely used to model homogeneous melting is employed[32]. In order to check whether this method overestimates $T_m$ or not, two-phase method calculations in *NPT* ensemble are also performed. In these simulations, the van der Waals functional vdW-DF is also used to justify the performance of PBE functional for dense atomic hydrogen. The facts of the perfect agreement of the $T_m$ predicted by these two functionals, and the good performance of van der Waals functional in describing $H_2$ dissociation by comparison with QMC[21] and dynamic compression experiment[22], give a strong support to the reliability of our DFT results. In fact, the validity of PBE functional in metallic hydrogen had been verified by method beyond DFT, such as the coupled electron-ion Monte Carlo simulations (CEIMC)[33]. The timestep of AIMD simulation is set as 0.5 fs. All simulated systems were observed to equilibrate rapidly within 0.5 ps. Moreover, all AIMD simulations run for at least 4 ps to collect enough data for statistics analysis. Other details of the simulations are given in the supplementary information.

**Nuclear quantum effects.** NQE influence is modelled using path integral AIMD simulations, in which a homemade code[19] is used. Due to the possible ambiguity of two-phase method in *NVT* ensemble[2], and considering the fact that there are no path integral algorithms for two-phase method in *NPT* or Z-method in *NVE* ensemble available currently, we did not calculate the NQE on melting directly, but turned to compute the NQE on superheating limit using a *heat-until-melting* approach. Since both $T_m$ and the superheating limit are at the same temperature level, the NQE corrections on them should be similar. The careful two-phase calculations as reported in Ref. [2] lend us the confidence of the validity of this treatment in dense hydrogen. Considering that the relevant temperature scale studied in this work is high enough, we employed 8 beads to discretize the integral path. The convergence quality is checked by using 32 beads, which do not change the results substantially.

## Acknowledgements

This work was supported by the National Natural Science Foundation of China under Grant Nos. 11274281 and 11672274, and by the CAEP Research Projects under Grant No. 2012A0101001 and No. 2015B0101005. Part of computation was performed on CCMS of the Institute for Materials Research at Tohoku University, Japan.

## Author contributions

H.Y.G conceived the project and performed the calculations. H.Y.G. and Q.W. analyzed the data and wrote the manuscript.

## Competing financial interests

The authors declare no competing financial interests.

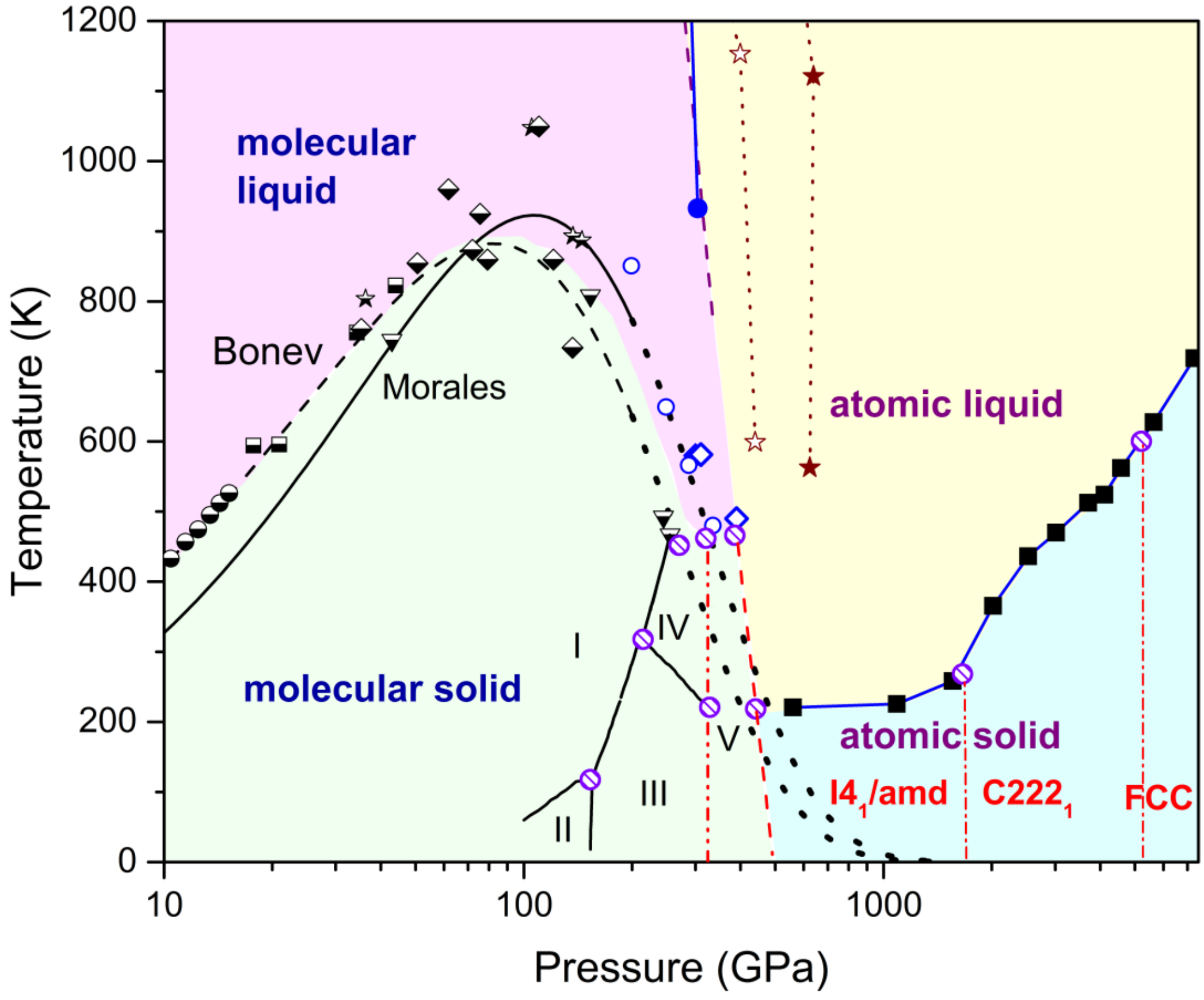

**Figure 1 | Phase diagram of dense hydrogen**. Theoretical melting points: the filled squares above 1.5 TPa are obtained in this work, those between 500 GPa to 1.5 TPa are taken from Ref. [2]; rhombuses—diatomic molecular *Cmca*-4 (filled—Liu *et al.*[29], opened—from Ref. [2]), open circles—Belonoshko *et al.*[28]; solid line—Morales *et al.*[33], dashed line—Bonev *et al.*[15], and dotted lines—the extrapolation of them using Kechin's equation. Experimental data on $H_2$ melting: half-filled symbols (circles—Datchi *et al.*[35], squares—Gregoryanz *et al.*[27], stars—Eremets *et al.*[25], rhombuses—Subramanian *et al.*[26], triangles—Howie *et al.*[11]). Experimental data on $H_2$ dissociation: blue filled circles—Knudson *et al.*[22] Dashed lines are theoretical $H_2$ dissociation, for which the liquid part is from PIMD+vdW-DF2 data of Morales *et al.*[30], and the solid part is a coarse estimation by using AIMD and ground-state DFT results, and is a guide to the eye only; the star-dotted lines indicate the dissociation region estimated by VMC-MD[18]. Solid molecular phase boundaries (I, II, III, and IV) are from Refs. [34,11], the boundary of phase V (vertical dash-dot-dot line) is inferred from Ref. [6] and is extended to low temperatures, and those of atomic solid (dash-dotted vertical lines) are guides to the eye only. Hatched circles denote possible triple points.

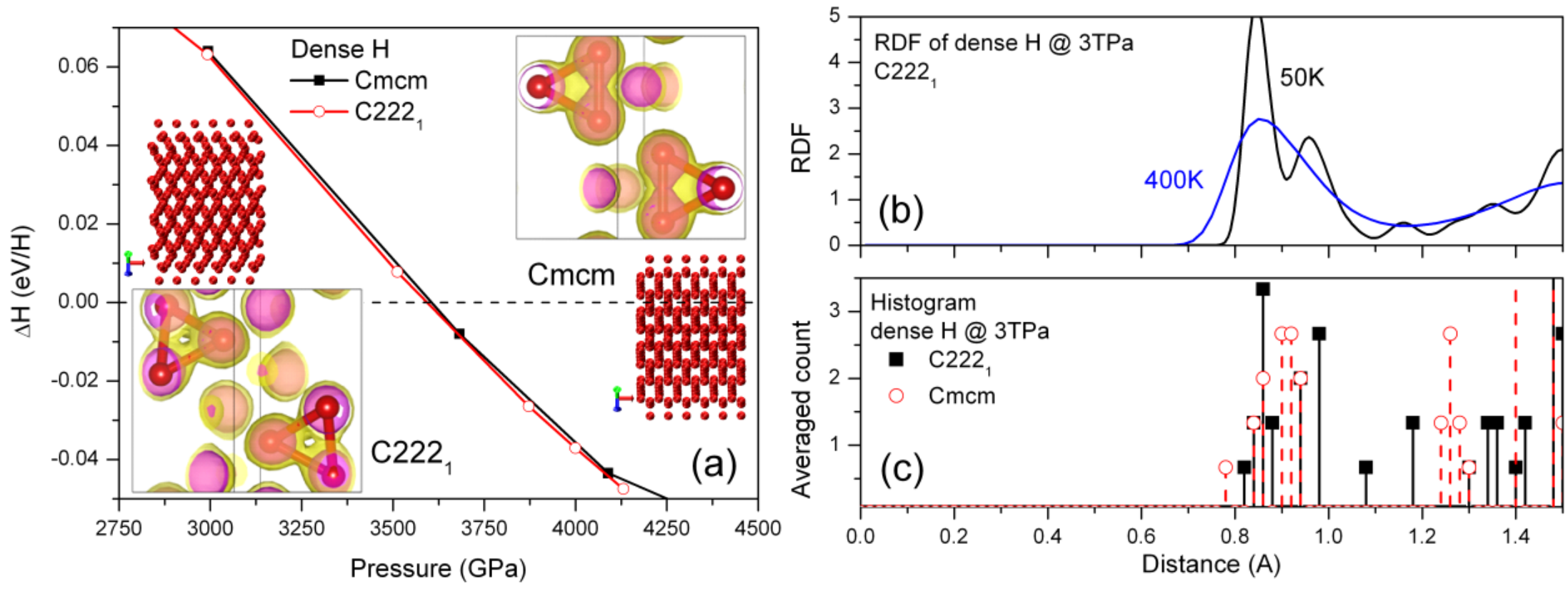

**Figure 2 | New candidate phase *C222₁* in dense hydrogen.** (**a**) Structure and electron localization function (ELF) in *Cmcm* and *C222$_1$* phases at 3 TPa, and their enthalpy difference with respect to a predefined equation of state, (**b**) RDF of *C222$_1$* at finite temperatures, (**c**) inter-proton separation histogram for the static lattice of *C222$_1$* and *Cmcm* structures.

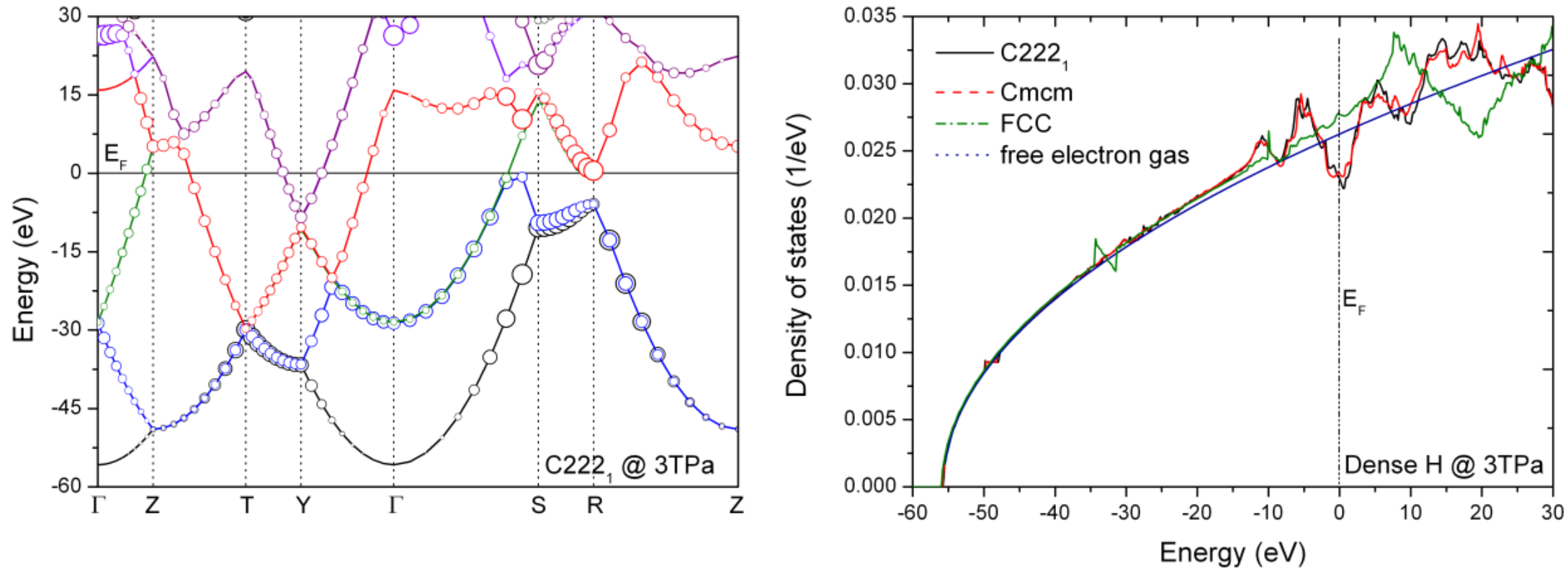

**Figure 3 | Nearly free electronic structure in dense hydrogen.** (**left**) Band structure of *C222₁* phase at 3 TPa, the size of circles is proportional to the *p* character of the wave functions, and (**right**) the density of states per electron for *C222₁*, *Cmcm*, and FCC phases of dense hydrogen at 3 TPa, and that of free electron gas, respectively.

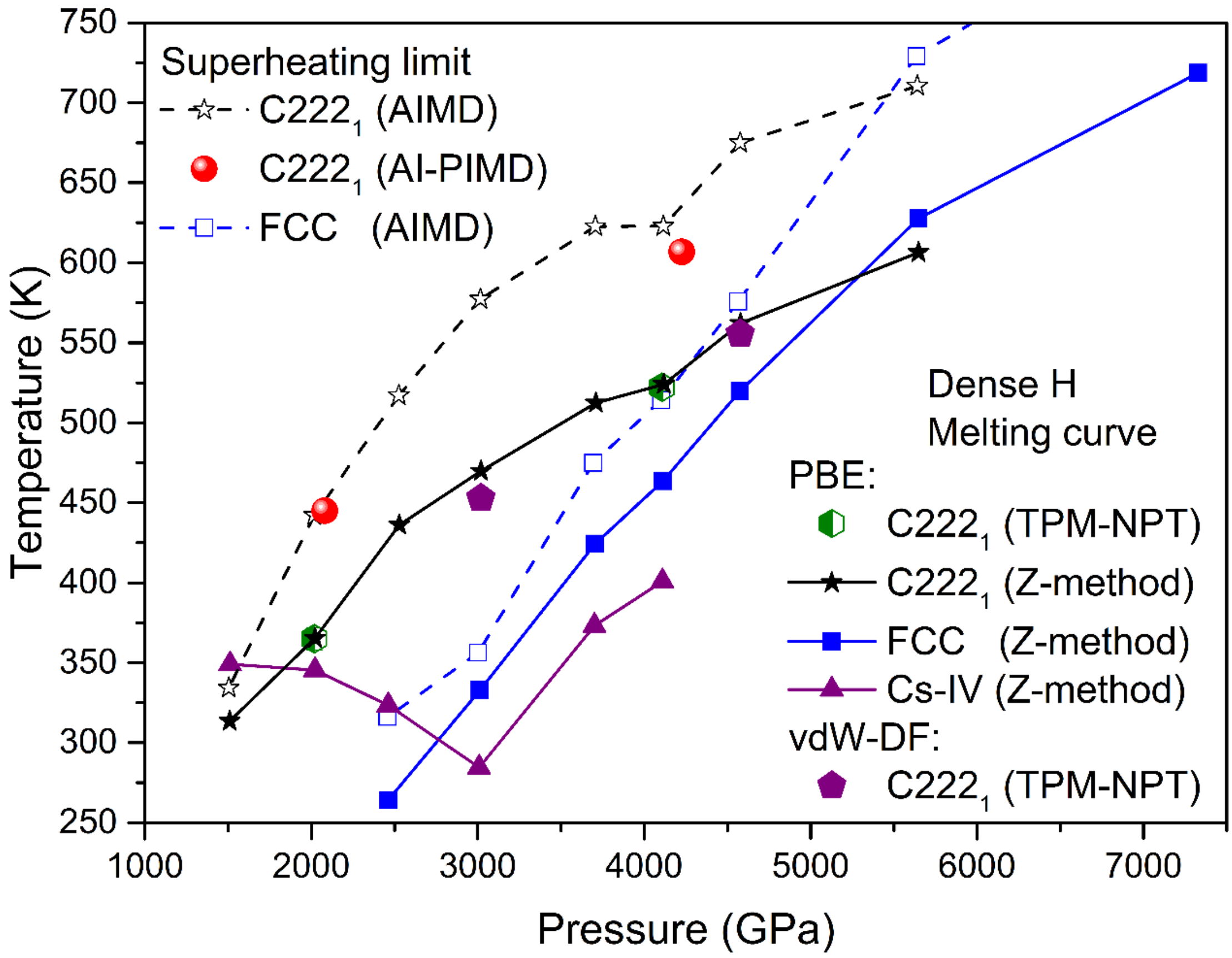

**Figure 4 | uprising of the melting temperature in dense hydrogen.** The calculated melting curves (for the *C222₁*, FCC, and Cs-IV phases, respectively) and superheating limit (only for the *C222₁* and FCC phases) of dense hydrogen from 1.5 to 7.3 TPa in several candidate solid phases by different *ab initio* simulation methods, which indicates the progressive enhancement of the stability of the solid phases of atomic hydrogen by compression. The previously speculated melting temperature is far lower and approaching 0 K.

# Predicted reentrant melting of dense hydrogen at ultra-high pressures

Hua Y. Geng* and Q. Wu

*National Key Laboratory of Shock Wave and Detonation Physics, Institute of Fluid Physics, CAEP; P.O.Box 919-102, Mianyang, Sichuan, P. R. China, 621900*

## Supplementary Information

### A. Compression effects on nearest-neighbor distance and pseudopotential

Usually compression reduces interatomic distance, and ultra-high compression might cause core overlap and leads to failure of the pseudopotential. However, the peculiarity of dense hydrogen is that the situation is just the opposite: the external pressure in fact compresses the inter-molecular space (or the van der Waals space) rather than the intra-molecular space, and the resultant dissociation of $H_2$ units leads to a larger nearest-neighbor (NN) distance, instead of the usually expected shortening of the NN distance. This unusual behavior of dense hydrogen has well been documented in Refs. [1,2]. In particular, the variation of the NN distance of monatomic hydrogen at ultra-high pressure regime has explicitly been given in the Fig. 14 of Ref. [2]. At higher pressures, for example the highest pressure of 7.3 TPa we have considered in this paper, the NN distances for the competing ground state of FCC and BCC phases are 0.78 and 0.76 Å, respectively. Both are slightly larger than the $H_2$ bond length of 0.74 Å at ambient conditions. This exceptional phenomenon guarantees that whenever a pseudopotential can be applied to low-pressure hydrogen, it also can safely be applied to dense hydrogen up to terapascal pressures. Figure 15 in Ref. [2] unambiguously demonstrated this by comparing the PAW results (from VASP) with the all-electron FLAPW results (from WIEN2k).

### B. Size-effect and k-point convergence

It is well known that MD simulations with a small cell size would suffer from finite size effects, and might predict a higher melting temperature if the Z-method is employed. For dense hydrogen, there are a lot of MD simulations having been performed and reported in literature, which have established a good database for a reliable estimate of the size effects in dense hydrogen. In the case of melting, Refs. [3-5] used 216, 432, and 1920 hydrogen atoms in the MD simulations, respectively. The perfect agreement of their obtained melting temperatures clearly indicates that the size effects are negligible when the simulation cell contains more than 200H atoms. This observation was further confirmed by detailed analysis given in Ref. [6]. Our simulation cells always contain 432-500H atoms, depending on the structure. They are large enough to eliminate possible size-effects. The perfect match of the melting temperatures calculated by the Z-method and the two-phase method justified this argument. Usually the former is much more prone to finite size effects and the latter is less sensitive to the cell size, as revealed by plentiful empirical experiences. On the other hand, our previous simulations showed that the convergence of the k-points is much more important than the size-effects for the cell size we used[4]. Figure S1 demonstrates the influence of different size of the k-point mesh on the melting temperature of Z-method. A good convergence is achieved with a 3×3×3 mesh, with an error of about 10 K in the resultant melting temperature.

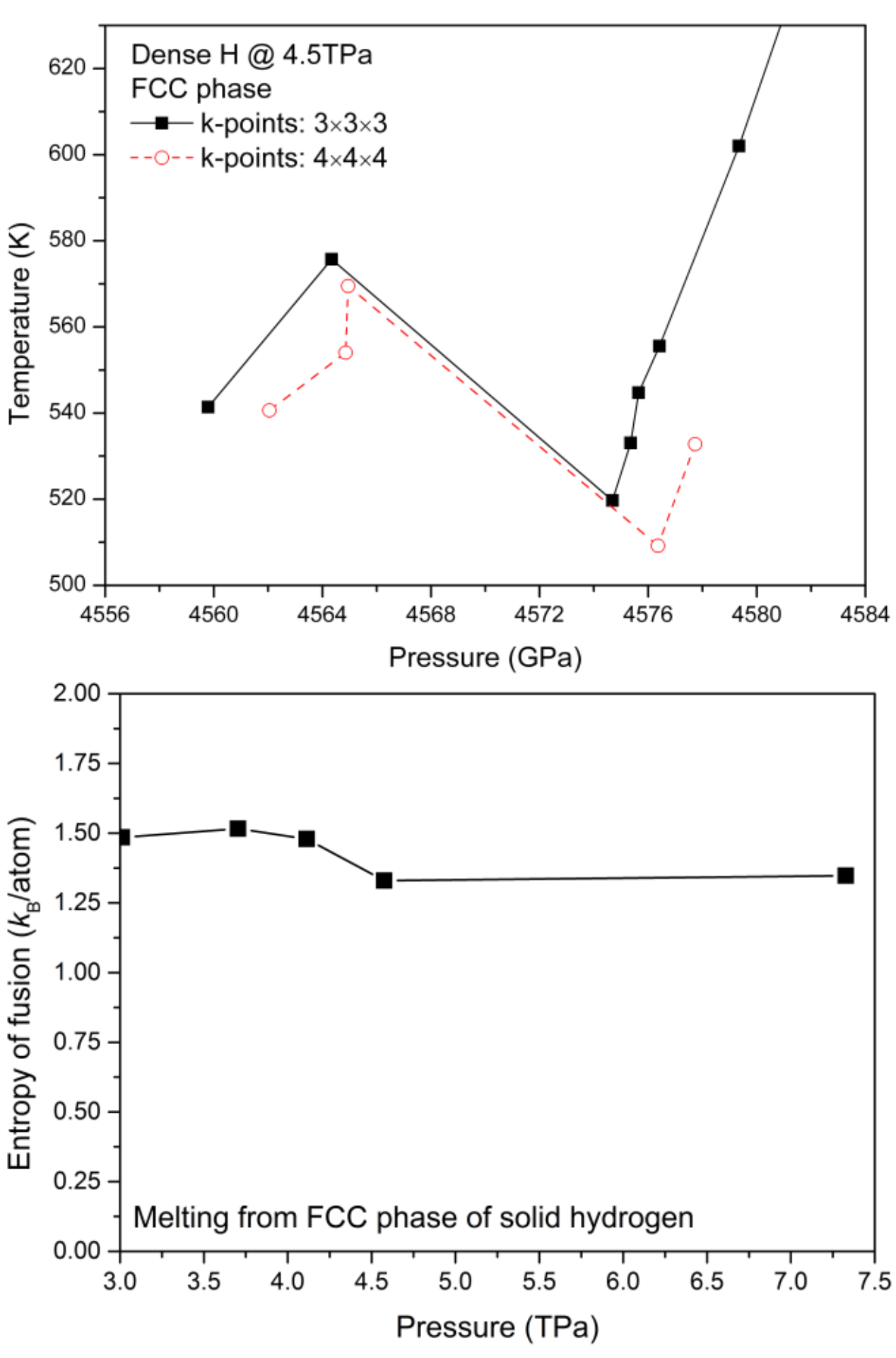

**Figure S1 | (top) Typical Z-curves in the Z-method for melting modeling.** The dense

hydrogen is at about 4.5 TPa heating from an FCC phase. The results of two different k-point meshes—3×3×3 and 4×4×4—are displayed. **(bottom) Entropy of fusion in dense hydrogen melting from FCC phase.**

## C. Melting in Z-method

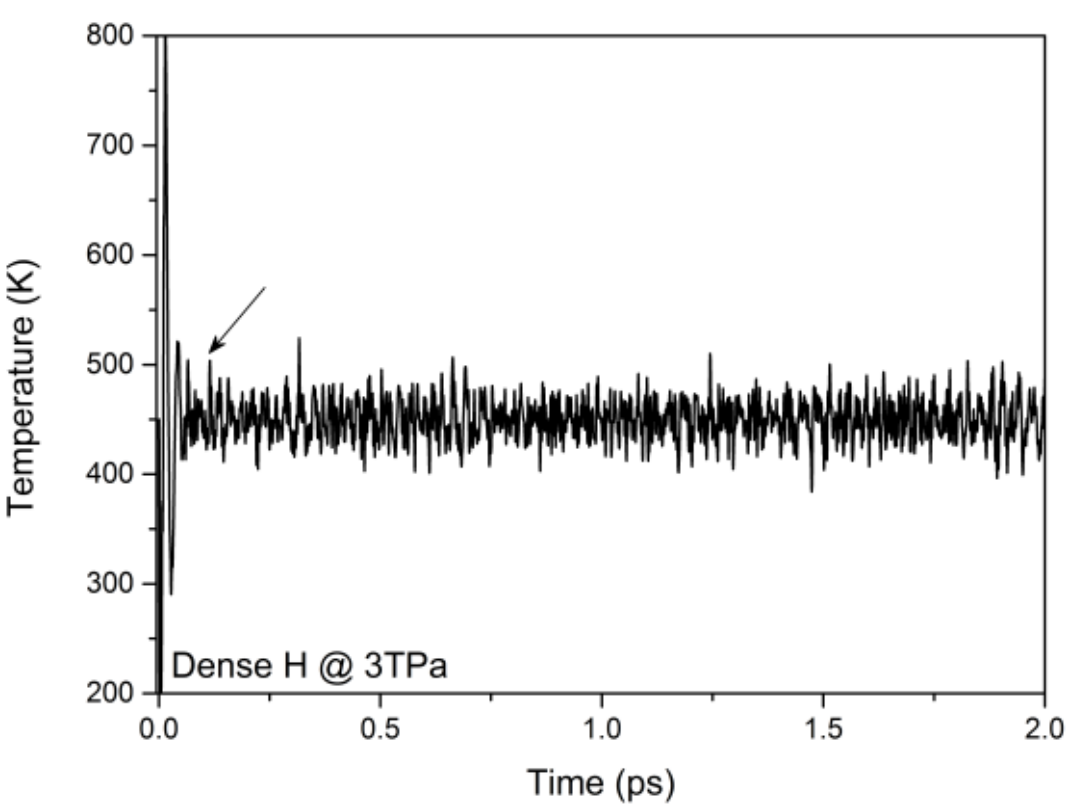

**Figure S2 | Typical equilibrating time-scale in AIMD simulations of dense hydrogen.** The structure is starting from the *Cmcm* phase at about 3 TPa, and equilibrates to the $C222_1$ phase at about 450 K with the Z-method. The arrow indicates the time when the system is fully equilibrated.

In AIMD simulations, to trigger a melting process is a rare event. It is thus crucial in Z-method that the system is fully equilibrated and is ergodicity before a melting point can be assessed. The equilibration in dense hydrogen is actually very fast. As shown in Fig. S2, the system equilibrates within 0.1 ps. This is due to the high vibrational frequency of hydrogen and the strong anharmonicity as revealed in Ref. [2]. The melting is initiated with a longer time, which requires about 0.25 ps (Fig. S3), but is much shorter than the typical time length of AIMD simulations we performed. The melting process leads to a redistribution of the energy between the potential and kinetic components, as revealed by the jumps shown in the Fig. S3.

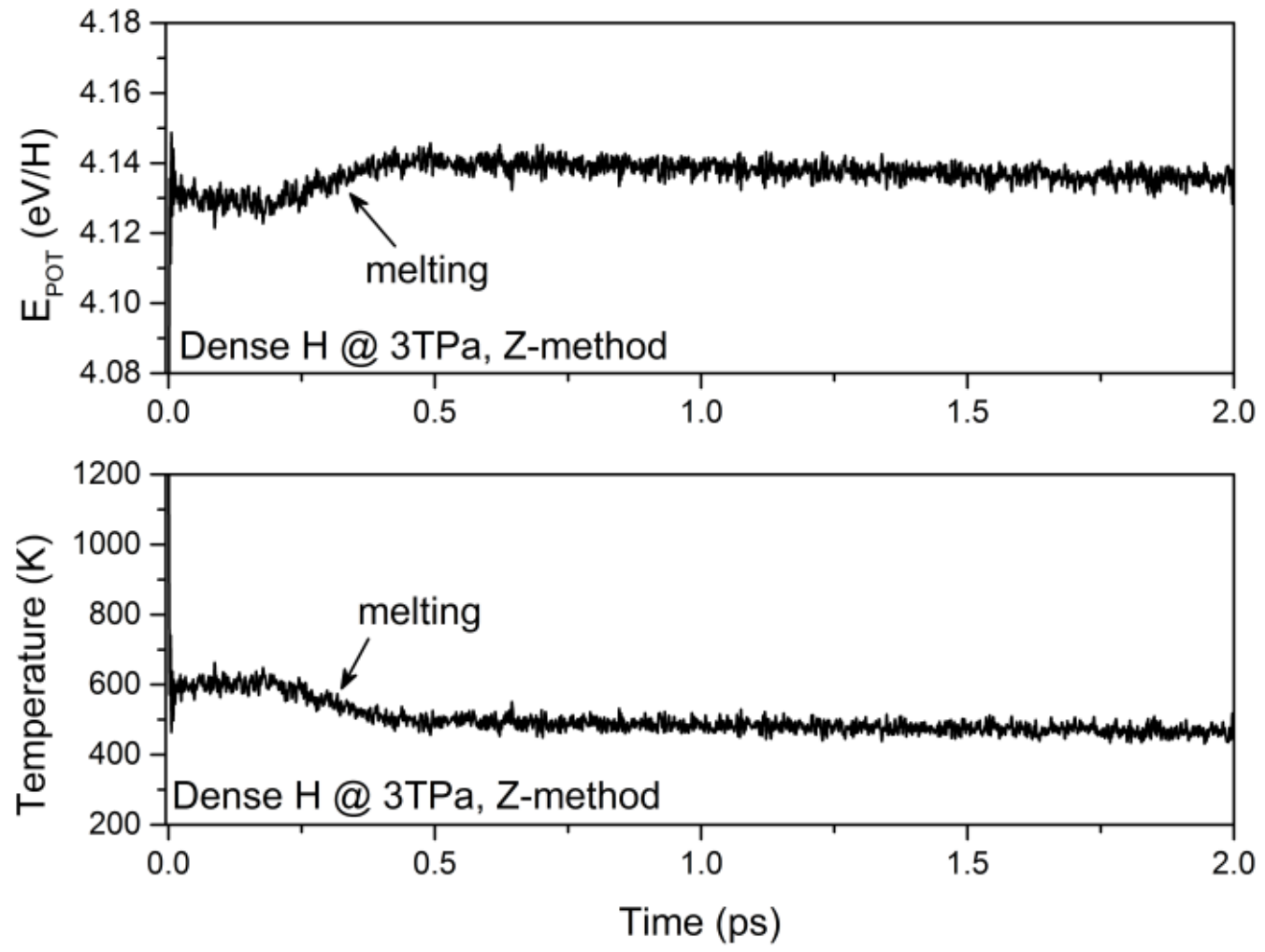

**Figure S3 | Evolution of the potential energy and temperature when the system melts in Z-method.** The structure is starting from the *Cmcm* phase at about 3 TPa, and equilibrates to the *C222₁* phase, then melts into the liquid phase. The arrow indicates the time when the melt occurs.

The variation of the mean squared displacement (MSD) and radius distribution function (RDF) of dense hydrogen at 3 TPa along the Z-curve before and after the kink are shown in Fig. S4. A typical liquid behavior is observed at 470 K, which is immediately after the kink where the lattice collapses. Previous investigation revealed that dense hydrogen might transform into a mobile anisotropic state that is not a real "liquid"[4]. Here we analyzed the particle distribution carefully, and failed to find any anisotropy. This confirms that the transition is indeed a melting.

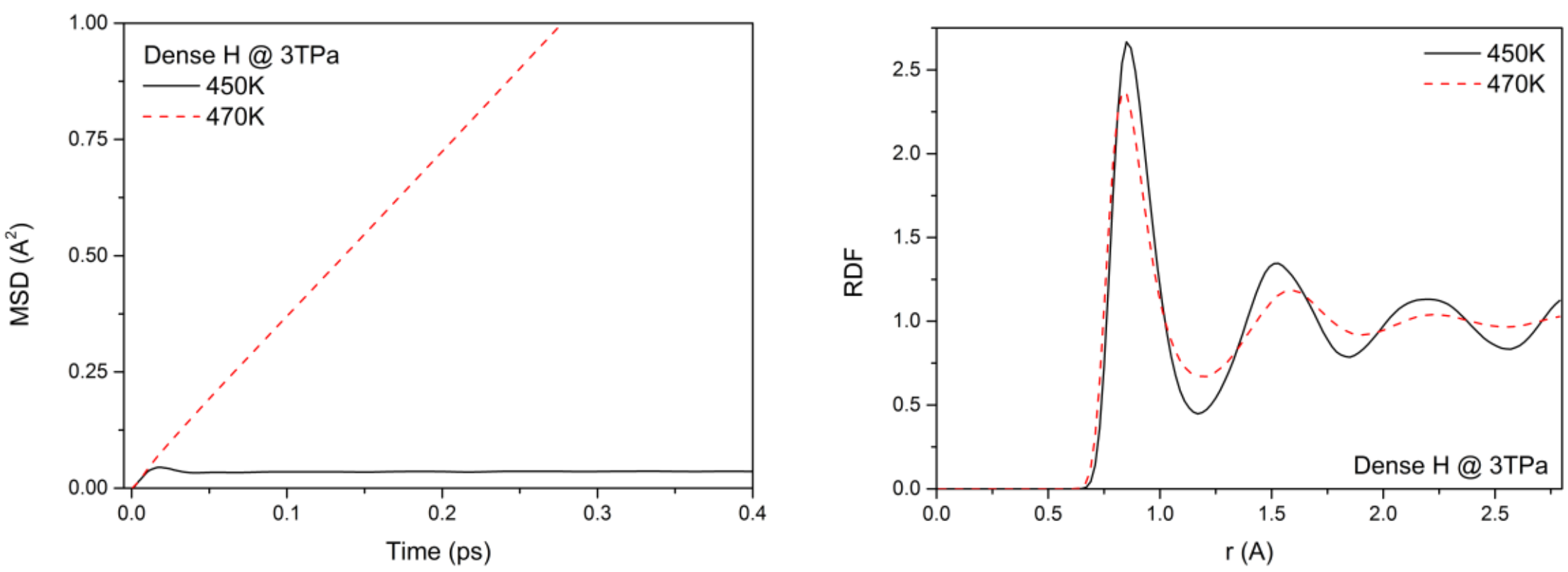

**Figure S4 | Evolution of MSD (left) and RDF (right) with temperature in Z-method when crossing the kinked Z-curve.** The structures are equilibrated before and after the occurring of the kink, respectively. The sharp variation of RDF and MSD from 450 to 470 K unambiguously indicates that it already melted at about 470 K.

## D. *NPT* ensemble and two-phase method

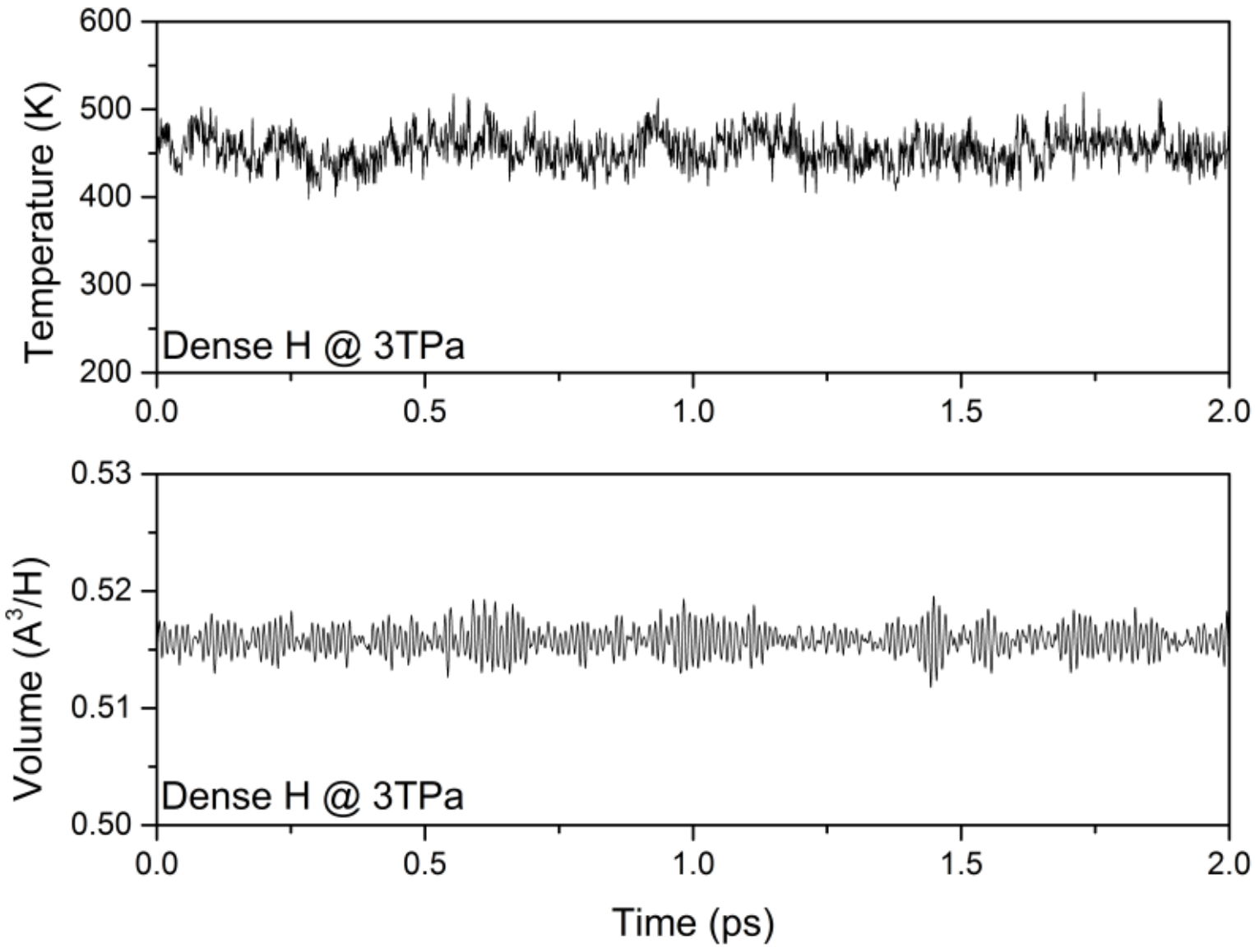

**Figure S5 | Fluctuations of the volume and temperature in an AIMD simulation for modelling an *NPT* ensemble.** The structure is equilibrated in the solid $C222_1$ phase at about 3 TPa. Efficient coupling with the thermostat and barostat is obvious.

The two-phase method in *NPT* ensemble is employed mainly for the purpose to check whether the Z-method overestimates the melting temperature or not. To achieve an *NPT* ensemble, the Parrinello-Rahman dynamics with Langevin thermostat as implemented in VASP is used. The fictitious mass and friction coefficient for lattice degrees of freedom are set to 20 atomic mass units and 55 $ps^{-1}$, respectively, and the friction coefficient for atomic degrees of freedom used in Langevin dynamics is set as 50 $ps^{-1}$. This setting leads to an efficient coupling with the thermostat and barostat, and equilibrates the system quickly, as indicated in Fig. S5. This good quality of AIMD simulations ensures a reliable two-phase modeling of the melting with *NPT* ensemble. A typical configuration of two-phase equilibration is illustrated by the snapshot shown in Fig. S6. It is necessary to point out that for the cell size we employed here, it is very difficult to maintain the two-phase equilibrating interface for a long time. Usually the system will evolve towards the solid or liquid phase quickly. Nonetheless, the uncertainty in the melting temperature introduced by this feature is less than 20 K, which is good enough for our purpose.

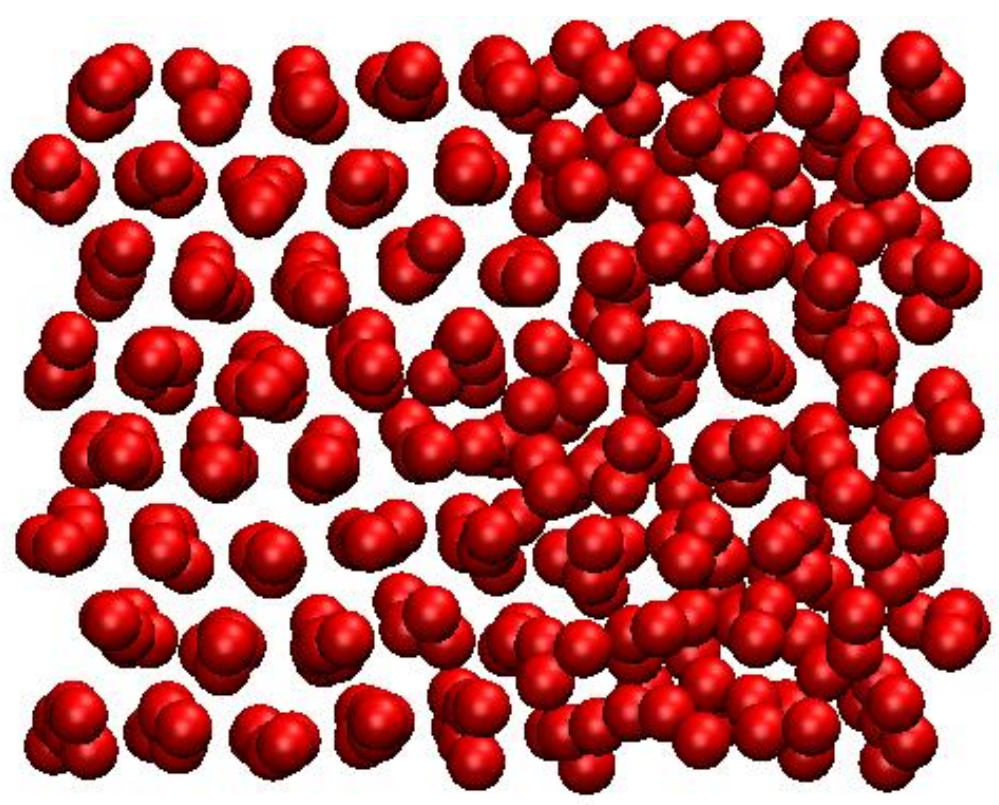

**Figure S6 | A snapshot of two-phase equilibrating.** The solid structure part is in the *C222*$_1$ phase, and the system is equilibrated at about 3 TPa and 460 K.

## Supplementary References

# Erratum: Predicted reentrant melting of dense hydrogen at ultra-high pressures [Sci. Rep. 6, 36745; doi: 10.1038/srep36745 (2016)]

Hua Y. Geng & Q. Wu

For the previously published article by Geng *et al.*[1], the authors have corrected the typographical errors about the labels of *Cmcm* and $C222_1$ phases. They all have 6 atoms in the primitive cell, instead of 8. The corrected label for them are *Cmcm*-6 and $C222_1$-6 (abbreviated as *Cmcm* and $C222_1$ for short), respectively. Their structural information is also listed in Table I.

TABLE I. The structure parameters for *Cmcm* (also denoted as oC12 in [2]) and $C222_1$ phase. The length of lattice vectors is in a unit of Å.

| Space group<br>Pressure | Lattice parameters | Multiplicity | Coordinates | | |
|---|---|---|---|---|---|
| $C222_1$ | (2.7952, 0.9479, 2.3372) | 8H | 0.86418 | 0.12707 | 0.33189 |
| 3 TPa | (90.0°, 90.0°, 90.0°) | 4H | 0.39617 | 0.00000 | 0.00000 |
| *Cmcm* | (1.0618, 3.0452, 2.6398) | 4H | 0.00000 | 0.89778 | 0.25000 |
| 1.5 TPa | (90.0°, 90.0°, 90.0°) | 8H | 0.00000 | 0.36050 | 0.58603 |